\documentclass[twocolumn,showpacs,preprintnumbers,amsmath,amssymb]{revtex4}

\usepackage{graphicx}
\usepackage{dcolumn}
\usepackage{bm}
\usepackage{amssymb}
\usepackage{indentfirst}
\usepackage{psfig,color}
\usepackage{epsfig}
\usepackage{epsf}
\usepackage{graphicx}
\usepackage{slashbox}

\begin{document}

\title{Scalar hadrons in $AdS_{5} \times S^{5}$.}
\author{Alfredo Vega} \author{Ivan
Schmidt}\affiliation{Departamento de F\'\i sica y Centro de Estudios
Subat\'omicos,  Universidad T\'ecnica Federico Santa Mar\'\i a,
Casilla 110-V, Valpara\'\i so, Chile}

\begin{abstract}


A holographic model is presented, which allows to describe scalar
hadrons with an arbitrary number of constituents.

\end{abstract}


\maketitle

\section{Introduction.}


The recent realization of the holographic principle as an AdS/CFT
correspondence has provided  a new theoretical tool for dealing with
strong coupling gauge theories. To the traditional models based on
lattice calculations, quark potential models, bag models, and
others, now we have at our disposal phenomenological QCD models
based on this correspondence. Although QCD is certainly not a
conformal field theory, it approaches the conformal limit in the
ultraviolet, and therefore the holographic description should be
valid.
On the other hand, confinement can be simulated by imposing
appropriate boundary conditions in the holographic variable
$z$~\cite{PolchStrass}. In the Hard Wall model, a sharp cutoff is
imposed. Moreover, confinement can also be introduced by modifying
the metric in order to mimic a confinement potential\cite{Andreev},
in which case a soft cutoff is introduced through a dilaton
field~\cite{KKSS} (Soft Wall model).


The mass spectra predicted by the Hard Wall model has the form $M
\sim l$ at high angular momenta $l$, which differs from the
quadratic dependence M$^2 \sim l$, expected within Regge
phenomenology. The correct Regge behavior can be obtained by
introducing a dilaton, which allows to retain the conformal metric
and in practice introduces a soft cutoff that depends on the dilaton
$\varphi$. So the usual procedure is to begin with
\begin{equation}
 S= \int d^{4} x dz \sqrt{g} e^{- \varphi (z)} \mathcal{L} .
\end{equation}
The Hard Wall model corresponds to a constant dilaton in the region
z $\leq \Lambda_{QCD}^{-1}$, tending to infinity for $z >
\Lambda_{QCD}^{-1}$. The introduction of a soft cutoff avoids the
appearance of ambiguities in the choice of boundary conditions. A
convenient and usual choice is $\varphi(z) = A
z^2$~\cite{KKSS,BdT1}, which is consistent with the usual Regge
behavior.


These models have been used to describe spectra of glueballs
(scalars~\cite{BdT2,BoschiBraga,CFJN} and vectors~\cite{CFJN}),
mesons~\cite{KKSS,BdT1,BdT2,KMMW} and baryons~\cite{BdT2}, getting
satisfactory results, but the hadronic spectra should also contain
exotic hadrons, like tetraquarks, pentaquarks and hybrids, which in
general have not been included in holographic models. Here we will
consider a model which in principle is able to describe these
hadrons.



\section{The Holographic Model.}


We consider a scalar field with mass in an $AdS_{5} \times S^{5}$
space with a dilaton field, whose action is given by
\begin{equation}
 \label{Accion10D}
 S= \int d^{10} x \sqrt{G} e^{- \varphi (z)} \frac{1}{2} [ G^{MN} \partial_{M} \hat{\phi}
 \partial_{N} \hat{\phi} + m^{2} \hat{\phi}^{2} ] .
\end{equation}
Here the M,N indexes are divided in $\mu,\nu$, which correspond to
$AdS_{5}$ coordinates and run over 0 to 4, and i,j indexes that run
over 5 to 9, and which are the compact space $S^5$ coordinates. The
ten dimensional space considered is described by the metric
\begin{equation}
 \label{Metrica10D}
 ds^2 = \frac{R^2}{z^2} ( \eta_{\mu\nu} dx^{\mu} dx^{\nu} ) + R^{2} \Omega_{ij} dy^i dy^j  ,
\end{equation}
where $\eta_{\mu\nu} = diag (1,-1,-1,-1,-1)$, R is the radius for
the $AdS_5$ and $S^{5}$ space and $\Omega_{ij}$ is the metric for an unitary
sphere in five dimensions. Beginning with (\ref{Accion10D}), the
equation for $\hat{\phi} (x,y)$ is
\begin{equation}
 \label{KleinGordon10D}
 \frac{1}{\sqrt{G} e^{- \varphi (z)}} \frac{\partial}{\partial x^M} ( \sqrt{G}
 e^{- \varphi (z)} G^{MN} \frac{\partial}{\partial x^N} \hat{\phi} ) - m^2 \hat{\phi} = 0  .
\end{equation}


Using $\hat{\phi} (x,y) = \phi (x) Y (y)$ it is possible to get an
equation for the AdS part, namely
\begin{equation}
 \label{KleinGordonAdS5}
 \frac{1}{\sqrt{g_{AdS}} e^{- \varphi (z)}} \frac{\partial}{\partial x^{\mu}}
 ( \sqrt{g_{AdS}} e^{- \varphi (z)} g^{\mu\nu} \frac{\partial}{\partial x^{\nu}} {\phi} )
 - m^{2}_{5} {\phi} = 0  ,
\end{equation}
where we have used
\begin{equation}
 \label{m5}
 m^{2}_{5} = m^{2} + \lambda^{2}  ,
\end{equation}
which is the mass of the scalar field in $AdS_{5}$ and $\lambda^{2} R^{2}=\kappa(\kappa+4)$. Notice that the
reduction of $S^{5}$ induces an additional mass term that depends on
$\kappa$, which is an integer.


Let us consider a mode propagating in the bulk, with the form
\begin{equation}
 \label{Phi}
 \phi (x) = e^{- i P \cdot \chi} f(z)  ,
\end{equation}
where $\chi$ corresponds to coordinates in four dimensional space
and $P^{2} = M^{2}$. Using (\ref{Phi}) in (\ref{KleinGordonAdS5}) we
obtain an equation which can be used in order to obtain the hadronic
spectra,
\begin{equation}
 \label{Ecuacion}
 [ z^{2} \partial^{2}_{z} - ( (d-1) z + z^{2} \partial_{z} \varphi (z) )
 \partial_{z} + M^{2} z^{2} - m^{2}_{5} R^2 ] f(z) = 0,
\end{equation}
written in a general form for an AdS space with $( d + 1 )$
dimensions.


For d=4, the conformal dimension for scalar fields is
\begin{equation}
 \label{88}
\Delta = 2 + \sqrt{4 + m^{2}_{5} R^{2}}.
\end{equation}
in $AdS_5 \times S^5$. Notice that $m_5$ depends of $\kappa$ and m.
Let us see what happens in the QCD side, where the quark field
(denoted by Q) has dimension 3/2, gluons (denoted by G) have
dimension 2, and the operator that gives one unit of angular
momentum has dimension 1. In this case, all O operators that
describe hadrons have dimension
\begin{equation}
 \label{8}
[O]=\Delta_0+l
\end{equation}
where  $\Delta_0$  corresponds to the contribution of the
constituents to the dimension of the operator, and the derivatives,
each with dimension one, give a total contribution l to (\ref{8}).
The AdS / CFT correspondence tells us that operators with dimension
[O] in a conformal field theory, are related with fields with
dimension $\Delta$ in $AdS_5 \times S^5$. This in practice provides
a mapping between $\kappa$ and the angular momentum $l$, which is
just phenomenological.

Of all possible cases in (\ref{88}), only some of them will be
related to operators of the conformal theory. To be more precise,
only some values for $m_5^2 R^2$ will be of interest to us, and the
selection of these values is give by solving the equation formed
using (\ref{88}) and (\ref{8}),
\begin{equation}
2+\sqrt{4+m_5^2 R^2}=\Delta_0+l,
\end{equation}
and then
\begin{equation}
 \label{9}
m_5^2 R^2=(\Delta_0+l-4)(\Delta_0+l).
\end{equation}
In other words, only Klein - Gordon fields in $AdS_5 \times S^5$
that satisfy (\ref{9}) can be used to describe scalar operators for
the field theory.
With the previous value for $m_5^2 R^2$, equation (\ref{Ecuacion}) can be
expressed as
\begin{equation}
 [ z^{2} \partial^{2}_{z} - ( (d-1) z + z^{2} \partial_{z} \varphi (z) ) \partial_{z} \nonumber
\end{equation}
\begin{equation}
 \label{Ecuacion2}
 + M^{2} z^{2} - (\Delta_0+l-4)(l+\Delta_0) ] f(z) = 0  .
\end{equation}

Notice that for l=0 the conformal dimensions are related to the
number of valence constituents, then equation (\ref{Ecuacion2})
could be used to study different kind of scalar hadrons using
different values to $\Delta_0$, or in other words choosing different
values for $m_5^2 R^2$ according to (\ref{9}). Since the conformal
symmetry is actually broken, there is a running coupling
constant~\cite{Kehagias, Nojiri, Huang, KKSS} and then R can change
too, and therefore here we consider it a parameter, contributing to
changes in $\Delta$. Moreover, $m_5^2 R^2$ is finite, and then there
must exist a maximum value for delta, ie the number of the valence
constituent in hadrons is bound. Table 1 shows different values for
$\Delta_0$, including the number of quarks (antiquarks) and gluons
in a hadron described by AdS modes.



\begin{table}[h]
\begin{center}
\caption{Relation between constituents in a hadron and the conformal
dimension. We consider hadrons with n quarks (and/or antiquarks) and
m gluons.\\}
\begin{tabular}{ c c c | c c c }
  \hline
  \hline
  & $\Delta_0$ & & & (nQ)(mG) & \\
  \hline
  & 3 & & & (2Q) & \\
  & 4 & & & (2G) & \\
  & 5 & & & (2Q)(1G) & \\
  & 6 & & & (4Q) & \\
  & 7 & & & (2Q)(2G) & \\
  & 8 & & & (4G) ; (4Q)(1G) & \\
  & 9 & & & (6Q) ; (2Q)(3G) & \\
  & 10 & & & (4Q)(2G) & \\
  \hline
  \hline
\end{tabular}
\end{center}
\end{table}


With this framework it is possible to study the spectroscopy
obtained using Hard Wall and Soft Wall models applied to hadrons.

\section{Spectra of scalar hadrons in a Hard Wall model. }


\subsection{Hard Wall model}
In this case equation (\ref{Ecuacion2}) can be expressed as
\begin{equation}
 \label{EcuacionHardWall}
 [ z^{2} \partial^{2}_{z} - 3 z \partial_{z} + M^{2} z^{2} - (\Delta_0+l-4)(l+\Delta_0) ] f(z) = 0  ,
\end{equation}
the normalizable solutions are
\begin{equation}
 \label{SlnHardWall}
 f(z) = z^2 J_{\alpha}(M z)  ,
\end{equation}
where $\alpha = \sqrt{4 + (\Delta_0+l-4)(l+\Delta_0)}$.


The values for M are obtained using the condition $f(z_0)=0$, where
$z_0 = \frac{1}{\Lambda_{QCD}}$. Thus, if the $n$-th zero of
$J_{\alpha} (z)$ is $\beta_{\alpha,n}$, the mass is \cite{BdT2}
\begin{equation}
 \label{MasaHardWall}
 M_{\alpha,n} = \beta_{\alpha,n} \Lambda_{QCD}^{(n)},
\end{equation}
where the index $n$ is related to the different radial excitations.


In Hard Wall models $\Lambda_{QCD}$ usually depends on the index $n$
(\ref{MasaHardWall}), because otherwise radial excitations are badly
described~\cite{BdT2}, and therefore it is necessary to consider a
different $\Lambda_{QCD}$ for each level. This can be corrected by
demanding good Regge behavior for radial excitations, as we will see
below.


For mesons we use $\Lambda_{QCD}=0.263$ GeV, and for glueballs this
parameter is fixed by considering that in the lattice the lightest
glueball has a mass of 1.61 GeV~\cite{MornPear}, which gives
$\Lambda_{QCD} = 0.313$ GeV for scalar glueballs. The results are
shown in Fig 1, where one can see that for small $l$, exempting the
pion, the agreement with data is quite good.
\begin{center}
\begin{figure}[h]
  \begin{tabular}{cc}
    \includegraphics[width=1.7in]{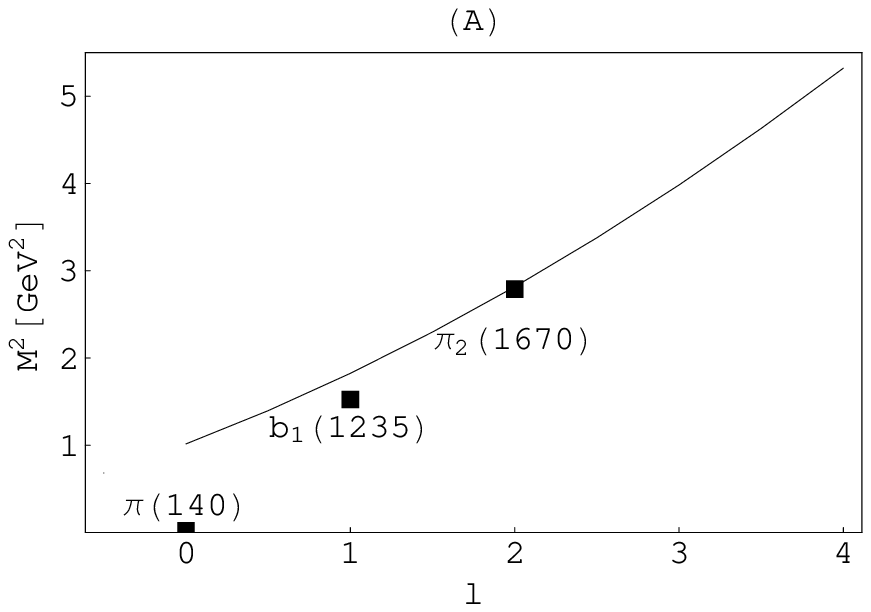} &
    \includegraphics[width=1.7in]{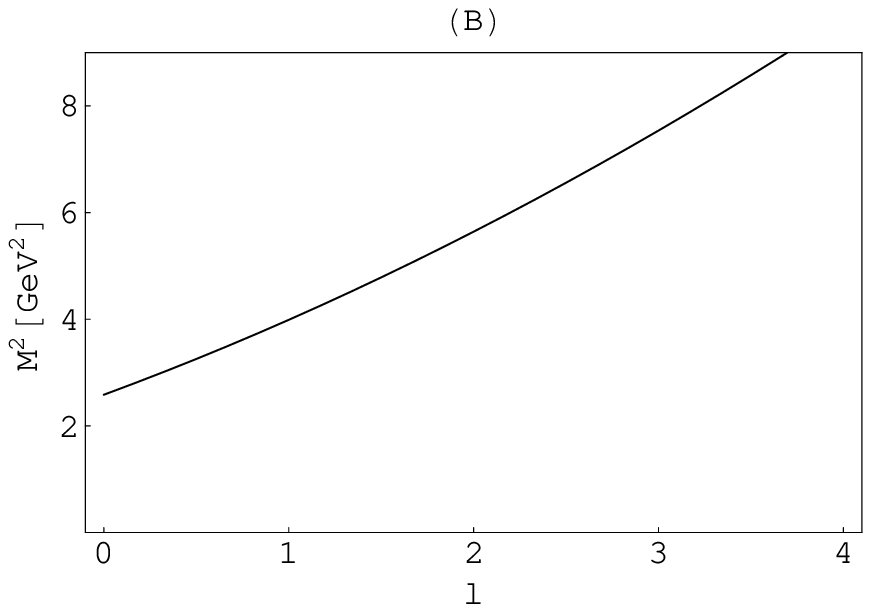}
  \end{tabular}
\caption{Spectra for states with different angular momenta. (A) mesons y (B) scalar glueballs.}
\end{figure}
\end{center}


Since there is no experimental data for exotic hadrons or hybrids,
it is not possible to fix $\Lambda_{QCD}$, and therefore we will use
both a minimum and maximum values for this parameter ( $0.1$ GeV and
$0.3$ GeV). Then we get a range of mass values for exotics and
hybrid hadrons, presented in Table 2 for $l=0$.
\begin{table}[h]
\begin{center}
\caption{Range of values for the mass in GeV of hadrons with n
quarks and/or antiquarks and m gluons. $M_{min}$ is calculated using
$\Lambda_{QCD}=0.1$ GeV and $M_{max}$ using $\Lambda_{QCD}=0.3$ GeV.
All cases considered here are for l $=$ 0.\\}
\begin{tabular}{ c c c | c c c | c c c }
  \hline
  \hline
  & (nQ)(mG) & & & $M_{min}$ & & & $M_{max}$ & \\
  \hline
  & (2Q)(1G) & & & 0.877 & & & 2.631 & \\
  & (4Q) & & & 0.994 & & & 2.981 & \\
  & (2Q)(2G) & & & 1.109 & & & 3.326 & \\
  & (4G) ; (4Q)(1G) & & & 1.223 & & & 3.668 & \\
  & (6Q) ; (2Q)(3G) & & & 1.335 & & & 4.006 & \\
  & (4Q)(2G) & & & 1.448 & & & 4.343 & \\
  \hline
  \hline
\end{tabular}
\end{center}
\end{table}

\subsection{Mesons in an improved Hard Wall model.}


Consider equation (\ref{Ecuacion2}) with $\Delta_0=3$.
\begin{equation}
[z^{2}\partial^{2}_{z}-(d-1)z\partial_{z}+z^{2}M^{2}-(l-1)(l+3)]f(z)
= 0  ,
\end{equation}
with normalizable modes which are given by
\begin{equation}
\label{FnOnda} f(z) = z^{2}J_{l+1}(z M)  .
\end{equation}


The mass spectra are obtained from $\Phi(x,z_{0} = 1 \diagup
\Lambda_{QCD}) = 0$, and therefore masses are determined by:
\begin{equation}
\label{masa} M_{1+l,n} = \beta_{1+l,n}\Lambda_{QCD}^{(n)}  ,
\end{equation}
where $\beta_{1+l,n}$ are zeroes of Bessel function that appear in
(\ref{FnOnda}), and $n = 1, 2, ...$.


A string mode with a node in the $z$ coordinate corresponds to
radial excitations, but as we just saw the Hard Wall model does not
give good results in this case without changing $\Lambda_{QCD}$ for
every level.


Nevertheless, values for $\Lambda_{QCD}^{(n)}$ can be obtained
beginning with $\Lambda_{QCD}^{(1)}$, by demanding good Regge
behavior. In fact,
Regge Trajectories as functions of a radial quantum number $n_{r}$
(related to $n$ by $n_{r} = n - 1$) can be described by

\begin{equation}
 \label{Regge}
 M^{2}(n_{r},l) = A n_{r} + M^{2}(0,l)  ,
\end{equation}
for fixed $l$.

\begin{figure}[h]
  \begin{tabular}{cc}
    \includegraphics[width=1.7in]{Apendice1.eps} &
    \includegraphics[width=1.7in]{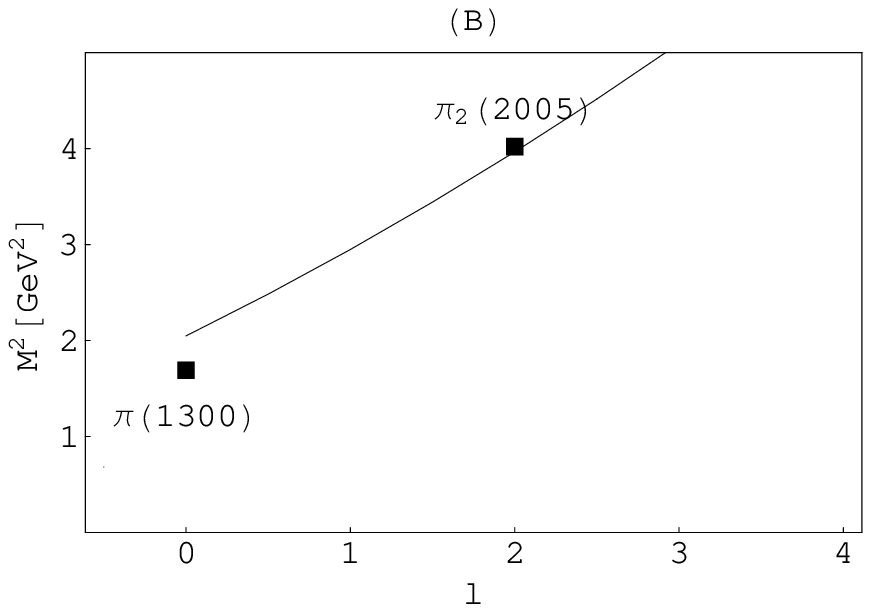}
  \end{tabular}
  \begin{tabular}{cc}
    \includegraphics[width=1.7in]{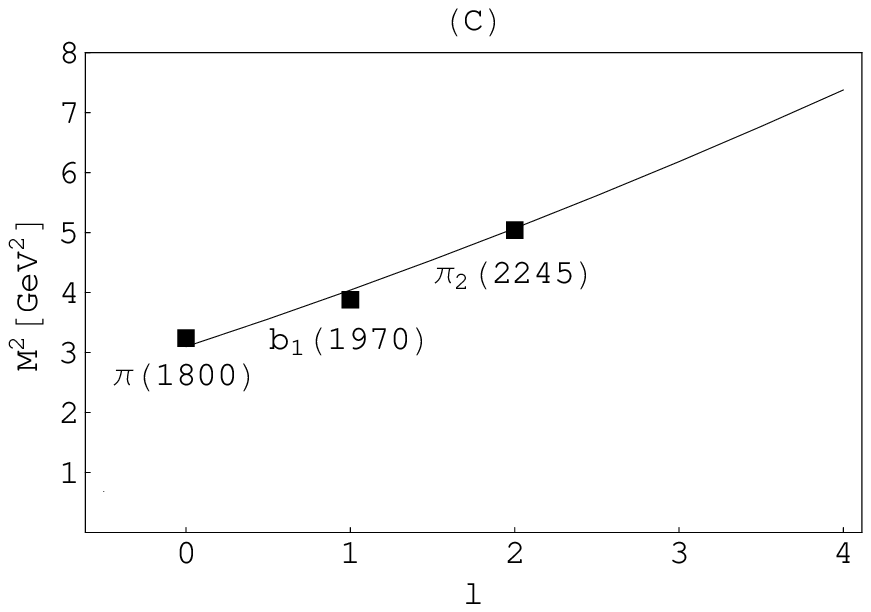} &
    \includegraphics[width=1.7in]{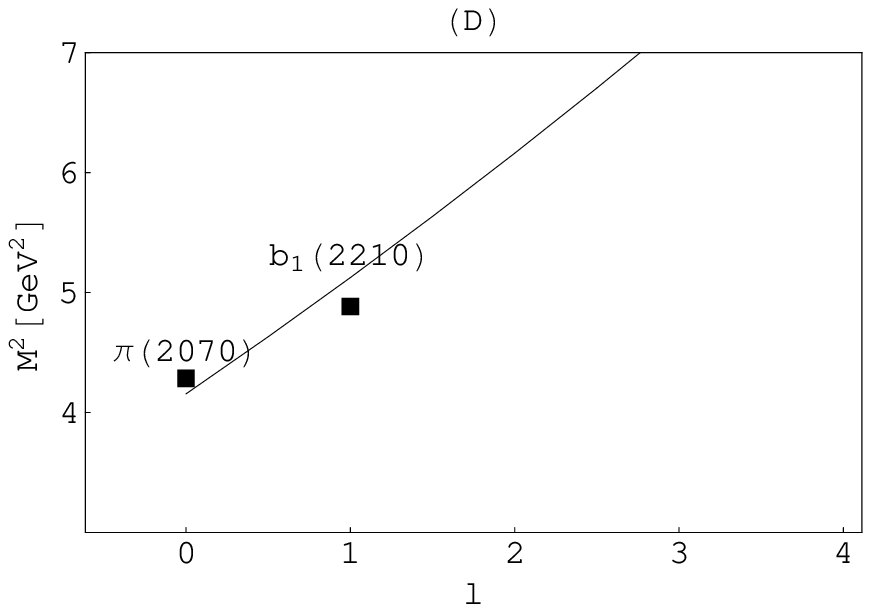}
  \end{tabular}
\caption{Spectra of scalar mesons, calculated within an improved
Hard Wall model. The figures correspond to different radial
excitations. (A) $n=1$ (B) $n=2$ (C) $n=3$ (D) and $n=4$. }
\end{figure}


An analysis of Regge trajectories for the radial excitations
$^{3}P_{2}, ^{3}P_{0}$ and $^{3}F_{2}$ shows that the slope is
constant, and its value is  $A = 1.04 \pm 0.01$ GeV \cite{AllgPeas}.


Using (\ref{masa}) and (\ref{Regge}), with $\Lambda_{QCD}^{(n)}
\rightarrow \Lambda_{n}$ we get

 \begin{equation}
 \label{Regge2}
 (\beta_{1 + l , n} \Lambda_{n})^{2} = A n_{r} + (\beta_{1 , 1} \Lambda_{1})^{2},
\end{equation}
and then

 \begin{equation}
 \label{LambdaN}
 \Lambda_{n} = \sqrt{\frac{A n_{r} + (\beta_{1 , 1} \Lambda_{1})^{2}}{(\beta_{1 + l , n})^{2}}}.
\end{equation}


Using $\Lambda_{1} = 0.263 GeV$, which is the value of
$\Lambda_{QCD}$ used in \cite{BdT2}, the other values for
$\Lambda_{n}$, calculated for $l = 0$, are

\begin{equation}
 \label{Lambda234}
 \Lambda_{2} = 0.204~ GeV,~~
 \Lambda_{3} = 0.173~ GeV,~~
 \Lambda_{4} = 0.153~ GeV.
\end{equation}


Using these values in (\ref{masa}) allows to obtain the scalar
mesons masses, which appear in Fig. 2, showing that for small values
of $l$ the improved model gives good results.

\section{Spectra of scalar hadrons in a Soft Wall model.}


Let us consider a Soft Wall model with a dilaton in its most general
form. In this case equation (\ref{Ecuacion}) can be written
as~\cite{HYY}
\begin{equation}
 \label{Ecuacion3}
 [ \partial^{2}_{z} - ( ( \partial_{z} C (z) ) \partial_{z} + ( M^{2} - \frac{m^{2}_{5} R^2}{z^2} ) ]
 f(z) = 0  ,
\end{equation}
where
\begin{equation}
 \label{B}
 C (z) = ( d - 1 ) \ln (z) + \phi (z)  .
\end{equation}


A correct choice of $\phi$ should be one which allows to get a
hadronic spectra consistent with Regge trajectories. A usual choice
for this function in $\phi (z) = A z^2$~\cite{KKSS,BdT1}, but a more
general choice is $\phi (z) = A z^2 + \delta \ln (z)$~\cite{HYY},
and then $C(z)$ has the form
\begin{equation}
 \label{B}
 C (z) = A z^2 + B \ln (z)  ,
\end{equation}
where $B = \delta + d - 1$.


Using
\begin{equation}
 \label{Transformador}
 f(z) = e^{\frac{1}{2} C(z)} \psi (z)  ,
\end{equation}
equation (\ref{Ecuacion3}) is transformed in a Schr\"{o}dinger type
equation
\begin{equation}
 \label{Sch}
 - \partial_{z}^{2} \psi + V(z) \psi = M^2 \psi  ,
\end{equation}
where
\begin{equation}
 \label{Potencial}
 V(z) = A ( B - 1 ) + \frac{B(B+2) + 4 m_{5}^{2} R^2}{4 z^2} + A^2 z^2
 .
\end{equation}
Where $m_{5}^{2} R^{2}$ is given by (\ref{9}). The potential in equation (\ref{Potencial})
should give a linear spectra in both $n$ and $l$, and for this
purpose taking $A$ constant is sufficient. With respect to $B$, it
has been shown that in order to get good Regge behavior, a
dependence on angular momenta is necessary~\cite{FBF}, a result
which is also present in our procedure.
A specific form of B, consistent with $M^2 \sim (n + l + v)$ for all
$\Delta_0$, is:
\begin{equation}
 \label{Parametrob}
 B = \frac{2 l + 3 l^2 + 4 \Delta_0 - 2 l \Delta_0 - \Delta_0^2 - 2 v + 8 l v + 4 v^2}{2 (l + v)}  .
\end{equation}


With this value for B we get a spectra given by:
\begin{equation}
 \label{M}
 M^2 = 4 A (n + l + v)  .
\end{equation}


In order to determinate $A$ and $v$, we use the $\pi$ and $\pi_2$
masses. With this we get $A = 0.346$ GeV$^2$ y $v = 0.0142$. Results
for mesons appear in Fig. 3.

\begin{figure}[h]
  \begin{tabular}{cc}
    \includegraphics[width=1.7in]{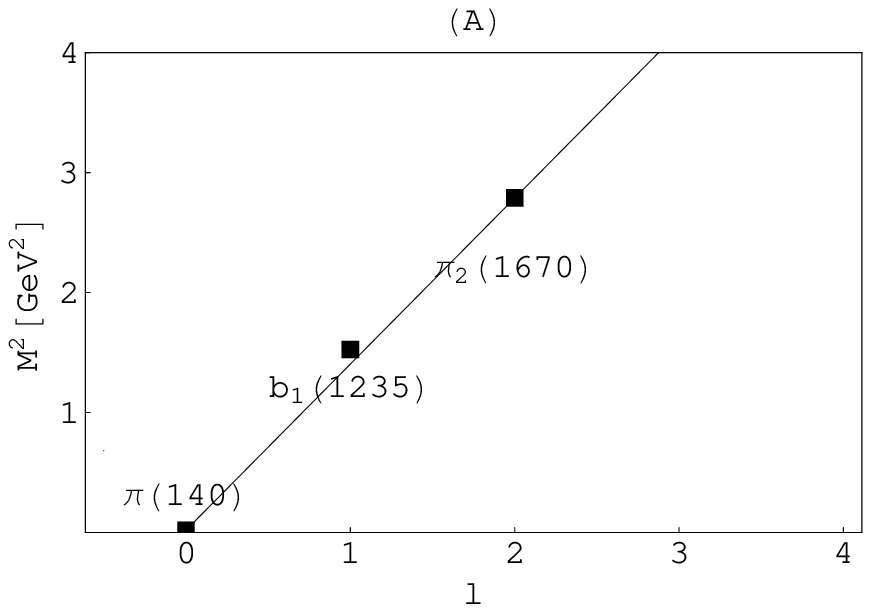} &
    \includegraphics[width=1.7in]{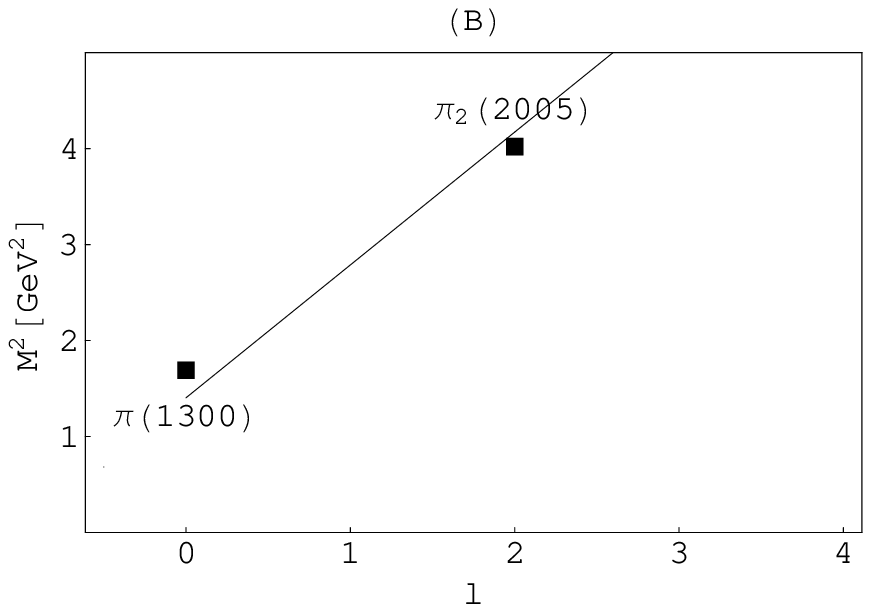}
  \end{tabular}
  \begin{tabular}{cc}
    \includegraphics[width=1.7in]{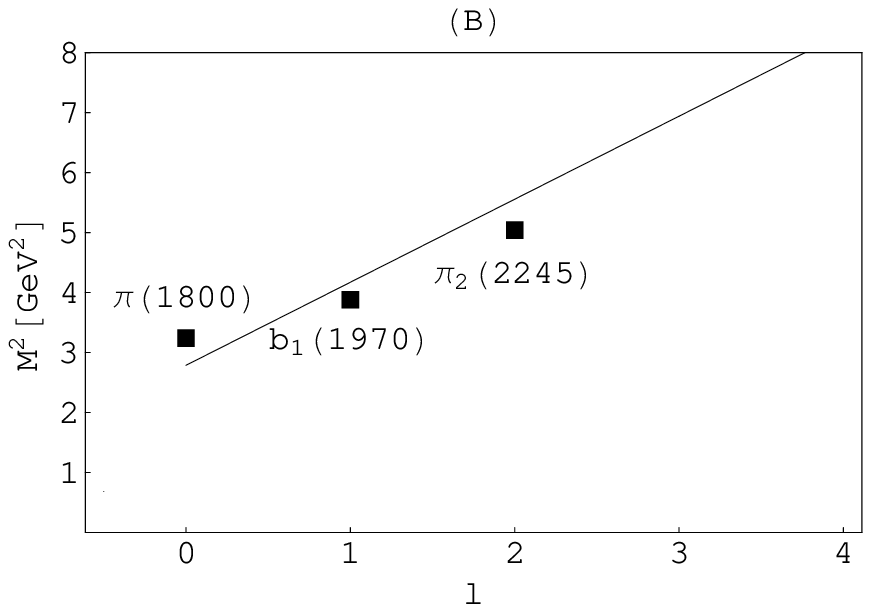} &
    \includegraphics[width=1.7in]{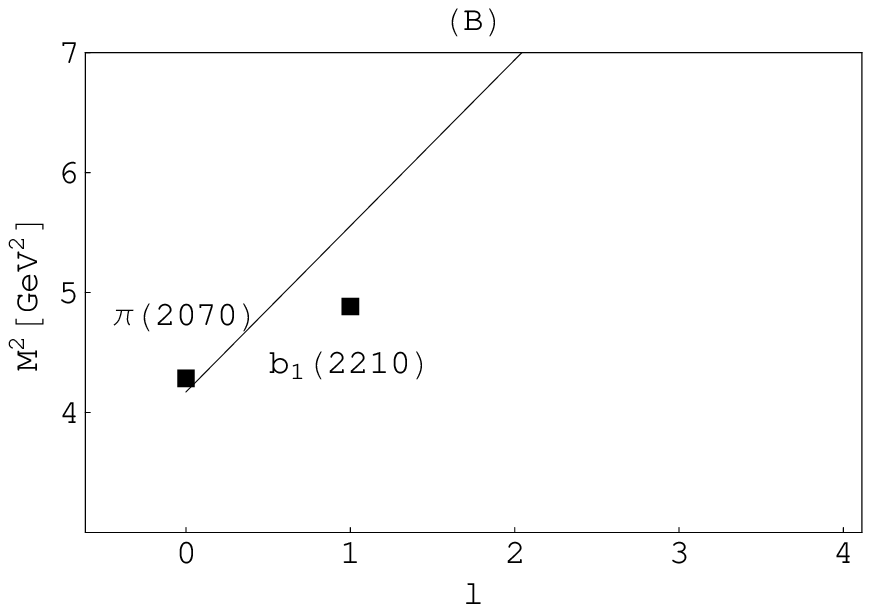}
  \end{tabular}
\caption{Scalar mesons spectra, calculated using a Soft Wall model with A $=$ 0.346 GeV
and v $=$ 0.0142. Differents graphics correspond to diferents radial excitations,
(A) n $=$ 0, (B) n $=$ 1, (C) n $=$ 3 y (D) n $=$ 3.}
\end{figure}


Considering that the slope in Regge trajectories is universal, the
previous $A$ value can also be used for glueball, and therefore we
only need to fix $v$. For this we use lattice values for the
lightest glueball~\cite{MornPear}, and find $v = 1.873$. Results for
glueballs appear in Table 3.
\begin{table}[h]
\begin{center}
\caption{Glueballs masses for different values of $n$ and $l$,
measured in GeV.\\}
\begin{tabular}{ c | c c c c c c c c c c c c c }
  \hline
  \hline
  \backslashbox{n}{l} & & & 0 & & & 1 & & & 2 & & & 3 & \\
  \hline
  0 & & & 1.61 & & & 1.99 & & & 2.32 & & & 2.60 & \\
  1 & & & 1.99 & & & 2.32 & & & 2.60 & & & 2.85 & \\
  2 & & & 2.32 & & & 2.60 & & & 2.85 & & & 3.08 & \\
  3 & & & 2.60 & & & 2.85 & & & 3.08 & & & 3.30 & \\
  \hline
  \hline
\end{tabular}
\end{center}
\end{table}


As can be seen, different values for $v$ must be given for each kind
of hadron considered, and therefore we did not make a table similar
to Table II in this case.

\section{Conclusions.}


The present work shows how holographic models can be applied to the
general problem of calculating hadronic spectra, considering even
exotic hadrons and hybrids. These were excluded in previous works
based on Hard and Soft Wall models, centered exclusively up to now
on glueballs, mesons and barions
\cite{KKSS,BdT1,BdT2,BoschiBraga,CFJN,KMMW}. 


In relation to the parameters of the dilaton field in a Soft Wall
model description, it is important that B depends on angular momenta
in order to get a good Regge behavior, which is already present in
(\ref{Parametrob}). This feature has also been mentioned in
\cite{FBF}.


At this moment, since the few parameters must be fixed by
experimental information,  it is not possible to make precise
predictions, but we can establish a range for the masses, depending
on a range for these parameters. In Hard Wall models the parameter
is $\Lambda_{QCD}$, for which a range of typical values is known.
This allows us to get a range for the possible masses. In principle
for Soft Wall models we can do the same, but the lack of knowledge
about the $v$ values does not allow to get a range for masses in
this case.

One interesting consequence of this approach is that there must
exist a maximum number for valence constituents inside hadrons.

\begin{acknowledgments}
We thank Stan Brodsky and Guy de T\'eramond for useful discussions.
The work of A.V. is supported by a Mecesup (Chile) scholarship.

\end{acknowledgments}


\end{document}